\documentclass[twocolumn]{webofc}

\usepackage[varg]{txfonts}   % Web of Conferences font
\usepackage{hyperref}
\usepackage{url}
\hypersetup{colorlinks=true,citecolor=blue,urlcolor=blue,linkcolor=blue}
\begin{document}
\title{Symmetry energy in dilute matter and the neutron skin
%and the PREX-CREX puzzle
}
%
% subtitle is optional

\author{\firstname{Panagiota} \lastname{Papakonstantinou}} %\inst{1}\fnsep\thanks{\email{ppapakon@ibs.re.kr}} 
        % etc.

\institute{Institute for Rare Isotope Science (IRIS), Institute for Basic Science(IBS), Daejeon 34000, Korea 
}

\abstract{Energy density functional (EDF) theory provides a unified framework for the description of nuclei and of infinite nuclear matter. In principle, it facilitates direct connections between nuclear data and the nuclear equation of state (EoS). Although in practice traditional nuclear EDF theory has strained to describe finite nuclei and infinite systems at the same time, recently developed extended EDF models overcome many of the limitations of traditional models in that respect. A recent challenge to EDF and EoS studies has come entirely from within nuclear structure, namely how to account both for the relatively thin neutron skin in $^{48}$Ca as extracted by the CREX experiment and the relatively thick neutron skin of $^{208}$Pb exctracted by the PREX-II experiment. The discrepancy suggests a surface and structure effect. The present study shows that the puzzle can be resolved in a non-relativistic framework by revisiting the nuclear surface tension and diffuseness, as driven in part by the EoS in dilute matter well below the saturation point and in part by the isovector gradient terms and spin-orbit potential. Such effects have no bearing on the EoS near and above saturation.  
}
\maketitle
\section{Introduction}
\label{intro}

It has proven a higly productive practice in nuclear physics to relate the properties of nuclear ground states and collective excitations to the equation of state (EoS) of infinite nuclear matter. 
%This is not unlike the way the density and the bulk and shear moduli in materials determine the properties and utility of the tools made of them. 
The most basic nuclear model, the liquid drop model, relates static nuclear properties with nuclear matter near its saturation density. The adoption of energy density functional (EDF) theory including not only number densities but also density gradients, current densities, and spin-orbit terms, has allowed for more sophisticated interpolations between the finite and infinite regimes. % with roughly ten phenomenological parameters. 
% ten parameters, one can describe quite accurately both finite nuclei and dense matter in the regime found in compact stars. 

The recent technological advances in rare-isotope physics and multi-messenger astronomy have been yielding a wealth of data and offer the promise to tightly constrain the nuclear equation of state from both the finite and the infinite side. Most relevant in what follows is the regime of cold nucleonic matter and therefore the discussion will be restricted to that. Specifically, the term EoS will be taken as a shorthand for the energy per nucleon of macroscopically homogeneous, uniform nucleonic matter as a function of the nucleon density $\rho$ and isospin asymmetry $\delta$, $E({\rho,\delta})$. Any parameters that characterize this density dependence will be referred to as {\em EoS parameters.}

The expansion of $E(\rho,0)$ in terms of the density provides the EoS parameters characterizing isospin-symmetric matter around its saturation density $\rho_0$, 
\[ 
E(\rho,0) = E_0 + \frac{1}{2} K_0 x^2 + \frac{1}{6}Q_0 x^3 + \ldots ; x\equiv (\rho -\rho_0)/3   . 
\] 
The expansion of $E(\rho,\delta)$ in terms of the asymmetry $\delta$ defines the symmetry energy 
$S(\rho) = \frac{1}{2} \left. \frac{\partial^2 E(\rho,\delta)}{\partial \delta^2} \right|_{\delta=0} $, 
whose expansion coefficients 
\[ 
 S(\rho) = J + Lx + \frac{1}{2} K_{\mathrm{sym}} x^2 + \frac{1}{6}Q_{\mathrm{sym}} x^3 + \ldots 
\]
characterize the EoS of asymmetric nuclear matter. 
Among the above, the lowest-order parameters $E_0, \rho_0, J$ are best constrained after decades of analyzing bulk and static nuclear properties. 
The parameters $K_0$ and $L$, both of which quantify the EoS's stiffness, have been constrained to a large extent with the help of collective excitations and properties of compact stars. See, {\em e.g.}, reviews \cite{Oertel2017,Roca2018,Garg2018}. EDF theory has been instrumental in this progress. 
Higher-order parameters have played minor roles until recently because they do not strongly affect basic observables. 

As more and more constraints emerged from heavy-ion collisions and astronomical observations, it was revealed that traditional EDF models which described nuclei accurately did not extrapolate well to infinite matter. Conversely, when trying to impose realistic EoS parameters on EDF models, results for nuclei deteriorated badly. The authors of ref. \cite{Roca2018} wrote somewhat fatalistically that ``going from successful models in the description of observables to the EoS is a well defined and safe strategy while ensuring reasonable parameters of the EoS do not necessarily lead to a good reproduction of the data on finite nuclei." The solution to this problem is so simple as to be invisible: Ensure that the EDF can accommodate the relevant EoS parameters independent of each other. Recent studies based on this strategy revealed the important role of the curvature of the symmetry energy $K_{\mathrm{sym}}$~\cite{ijmpe31,XuP2022}, whose value in traditional functionals is, unphysically, fixed though the lower-order parameters. They also revealed the importance of decoupling the modeling of the effective mass from the density dependence of the static functional~\cite{kids_nuclei1}. 

The role of $K_{\mathrm{sym}}$ in reconciling nuclei with dense matter is currently more and more recognized. 
%and a recent study exploited it to reconcile the thick neutron skin of $^{208}$Pb, which typically requires a lrage value of $L$, with astronomical observations which were not compatible with such high values of $L$. 
However, varying $K_{\mathrm{sym}}$ cannot resolve a very elementary puzzle, involving contradictory observations in the same density and asymmetry regime, namely the thick neutron skin of $^{208}$Pb~\cite{PREX} and the thin neutron skin of $^{48}$Ca~\cite{CREX}, or PREX-CREX puzzle. 

Next, we take the measurements from the PREX-II and CREX experiments at face value and, like many before, attempt to reconcile them, this time with promising results. The solution seems to lie in the density dependence of the EoS of dilute matter and at the same time structure effects on the nuclear surface, which in the non-relativistic framework used here are represented by the isovector gradient terms and the spin-orbit potential. 

\section{Survey of parameter space} 

A hint towards the present approach was obtained from surveying the parameter space of the EoS very widely: With extreme enough $Q_{\mathrm{sym}}$ and $R_{\mathrm{sym}}$ values, such that the symmetry energy was enhanced at low densities, reconciliation was achieved~\cite{rila2022}. In addition, the electric dipole polarizability of $^{208}$Pb was reproduced. Of course, the extreme values also resulted in undealistic dense EoS, but the main take-away of that survey was that the EoS of dilute matter could not be na\"ively extrapolated from the saturation point. The behaviors in the two regimes need to be decoupled. In the extended non-relativistic framework employed in this work, a simple way to modify the dilute-matter EoS without affecting the saturation and dense regimes is to introduce an additional density-dependent term to the functional that is only active at low densities. The latter feature can be ensured by an exponential decay and is explored presently. 
Although the precise physical meaning of such a phenomenological term is open to interpretation in terms of correlations or clustering--see also the discussion in \cite{Li2017}--its absence is not warranted either, because dilute gas-like matter and near-saturated matter are different enough to  be characterized by EoSs with very different polytropic laws and expansion coefficients.

We begin with a standard ``3+4" KIDS EoS and EDF~\cite{symmetry}, determined via three parameters $(E_0,\rho_0,K_0)$ for the symmetrinc-matter EoS and four parameters $(J,L,K_{\mathrm{sym}},Q_{\mathrm{sym}})$ for the symmetry energy, as well as the independently varied isoscalar effective mass $(m^{\star}/m)$, isovector enhancement factor $\kappa$, isoscalar and isovector gradient terms $C_{12},D_{12}$, and the spin-oprbit coupling $W_0$. 
Throughout this study, the symmetric-matter EoS parameters are fixed at $\rho_0=0.16$~fm$^{-3}$ and $(E_0,K_0)=(-16,240)$~MeV while the isoscalar effective mass and isovector enhancement factor are fixed at $(m^{\star}/m,\kappa)=(0.70.0.40)$. The isoscalar gradient term is also fixed at $C_{12}=-75$~MeV~fm$^5$--see \cite{symmetry} for conventions.
Results were obtained for all combinations of the following symmetry-energy parameter values: $J=30,31,32,33$~MeV, $L=40,45,\ldots,75$~MeV, $K_{\mathrm{sym}}=-200,-150,0$~MeV, and $Q_{\mathrm{sym}}=0,650,1200$~MeV and for all combinations of $D_{12}=-5,0,\ldots,25$ and $W_0=126,128,\ldots,140$~MeV~fm$^5$. 
As acceptable combinations of EoS and EDF parameters are considered only those that give an average accuracy of $1\%$ for a set of 17 data for energies and charge radii of closed-shell nuclei. 

\begin{figure}[h]
\includegraphics[width=7.5cm]{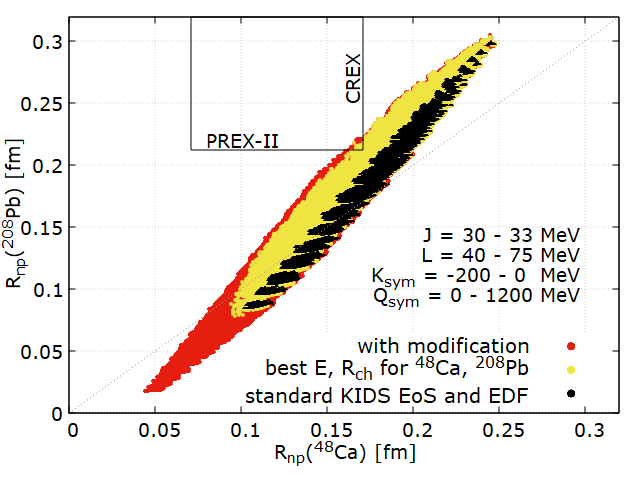}
\caption{Scatter plots of predictions for the neutron skin thickness of $^{48}$Ca and $^{208}$Pb 
for various EDF models which reproduce energies and charge radii of closed-shell nuclei within $1\%$ on average and with realistic EoS parameters. 
The rectangle encloses the CREX and PREX-II results with their reported errorbars. 
Black: Standard KIDS EoS and EDFs. Red: With modification of the dilute-matter EoS. 
Yellow: The subset for which the energies and charge radii of $^{48}$Ca and $^{208}$Pb are reproduced within $1\%$ each. 
\label{fig:skin} }
\end{figure}

The procedure produced numerous acceptable EDFs of different symmetry energy parameters. 
Their predictions for the neutron-skin thickness of $^{48}$Ca and $^{208}$Pb are shown in the form of a scatter plot of black points in figure~\ref{fig:skin}. 
The lines forming a rectangle enclose the overlap of the reported $1\sigma$ confidence interval from the CREX experiment, $0.121\pm 0.050$~fm, and the $1\sigma$ confidence interval from the PREX-II experiment, $0.283\pm 0.071$~fm. 
The black points form a somewhat broad structure that lies well outside the rectangle. 
The conclusion is that the standard KIDS EDF with up to 3+4  independent EoS parameters cannot account for both measurements at the same time, in line with previous studies employing traditional Skyrme functionals and relativistic mean-field theory. 

\begin{figure}[h]
\includegraphics[width=7cm]{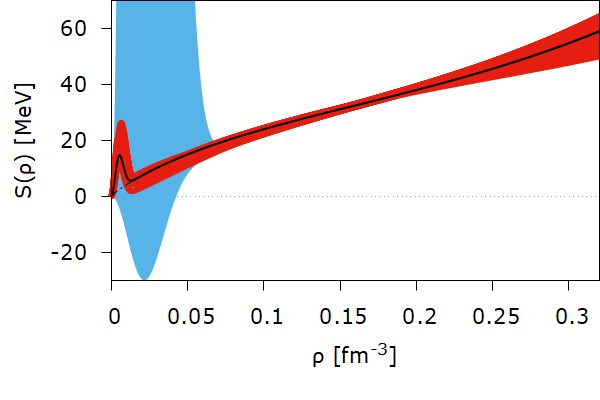}
\caption{Symmetry energy as a function of the density. Blue: The regime of dilute-matter EoS explored in this work. Red: EoSs which were found to reproduce both the CREX and PREX-II measurements as well as basic nuclear properties. Black: The example discussed in Sec.~\ref{sec:surf}. 
\label{fig:EoSs} }
\end{figure}

Next, a modification is introduced to the EoS of very dilute matter. 
For the purposes of the present work, the term added to the effective potential, in a notation similar to Skyrme and KIDS EDFs~\cite{symmetry}, is of the form 
\[ 
 \frac{1}{6} (t_d + y_dP_{\sigma})\rho^{a_d}e^{-b_d\rho^2} \delta (\vec{r}_1-\vec{r}_2) 
\]
which contributes to $S(\rho)$ a term equal to 
$-\frac{1}{48}(t_d+2y_d)\rho^{a_d+1}e^{-b_d\rho^2}$.
At present, we fix the power $a_d=1.0333$, one of the few values tried thus far with similar outcomes.  
The parameter space for $t_d,y_d,b_d$ is explored widely for each one of the same 
$(J,L,K_{\mathrm{sym}},Q_{\mathrm{sym}}, D_{12},W_0)$ values mentioned earlier. 
An illustration of the EoS regime the new term can span is shown in Fig.~\ref{fig:EoSs} in the form of the blue-shaded area. 
As before, the only acceptable combinations of parameters are those that give an average accuracy of $1\%$ for the same set of 17 data for energies and charge radii of closed-shell nuclei.

The procedure yielded a great many acceptable EDFs of different symmetry energy parameters including modification terms.  
Their predictions for the neutron-skin thickness of $^{48}$Ca and $^{208}$Pb are shown in the form of a scatter plot of red points in  fig.~\ref{fig:skin} (including the yellow and black areas). The red points form a much wider structure than the black ones and many of them are able to reproduce both the CREX and PREX-II results. 
As an additional refinement, the sets were selected for which the energies and charge radii of the two nuclei of interest are each reproduced within $1\%$. The results for the neutron skin thickness, shown in yellow, still span a large domain also overlapping with the CREX and PREX-II results. 
Inspecting the successful EoSs and EDFs, one finds that the reconciliation favors higher $(J,L,-K_{\mathrm{sym}},Q_{\mathrm{sym}})$ values among those surveyed. 
It also favors higher $D_{12}$ and $W_0$ values. 

The successful EoSs, which correspond to the yellow points lying within the rectangle in fig.~\ref{fig:skin}, are shown in fig.~\ref{fig:EoSs} in red color.  
As in the previous pilot study~\cite{rila2022}, an enhancement is observed of the symmetry energy at low densities. 
According to both studies, then, it is possible to describe both the CREX and PREX-II results by considering an enhanced symmetry energy in dilute matter which cannot be inferred from the EoS of dilute matter by simply using the low-order Taylor expansion coefficients. 

The physical explanation for the %origin of the 
enhancement at low densities is open for discussion~\cite{Li2017}. 
One possibility to consider is the modification of the EoS at low densities owing to the onset of a different phase such as clustering. 
Isoscaling experiments have deduced the symmetry energy at low densities, between roughly $0.002$ and $0.05$~fm$^{-3}$, to be of the order of $10$~MeV~\cite{Nat2010,Wada2012} and attribured its enhancement to clustering correlations. 
A complementary possibility requires us to think of the EoS inferred this way as an {\em effective EoS} for dilute matter on the surface of a nucleus, subject to the attrative field of the nucleus. 
The local minimum at $0.01-0.02$~fm$^{-3}$  
corresponds to an average closest-neighbor distance of, very roughly, $4-5$~fm, which is of the order of the size of a nucleus. The local maximum then in effect prevents the surface nucleons from drifting outwards to form lower-density regions.

\section{Surface pressure?\label{sec:surf}} 
 
The results so far point to the following mechanism for reconciling the CREX and PREX-II measurements within EDF theory: 
First, a somewhat stiff EoS has to be assumed in the saturation region, on the higher side of realistic $(J,L,-K_{\mathrm{sym}},Q_{\mathrm{sym}})$ values.  
Traditional functionals corresponding to that EoS, including a standard KIDS model, will predict  thick neutron skins for both nuclei---and for most nuclei in gerenal, according to known systematics.   
A subsequent modification of the model symmetry energy at low densities, with a balancing modification of the gradient and spin-orbit terms so as to maintain a realistic description of the energies and charge radii, seems to create enough pressure to reduce both neutron skins to their observed values. 
The effect is stronger for $^{48}$Ca.
Next, we will demostrate the mechanism with a numerical example. 
%A numerical example from the present study is shown in figure \ref{fig:example}. 

\begin{figure}[h]
\includegraphics[width=7.5cm]{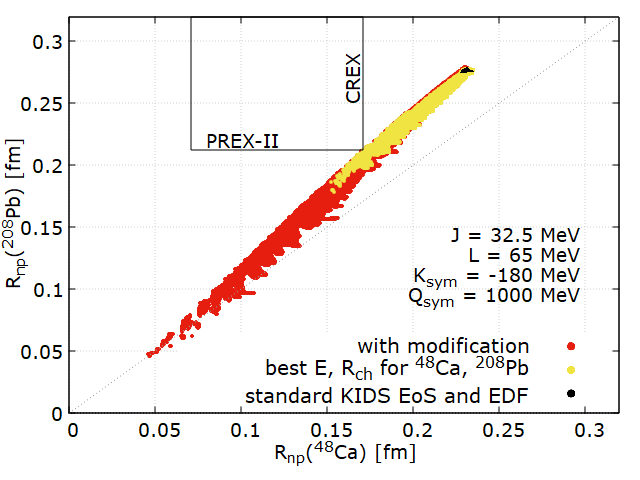}
\caption{Same as fig.~\ref{fig:skin} but only for the shown values of symmetry-energy parameters. 
\label{fig:jlkq} }
\end{figure}

The symmetry energy parameters are chosen to be $(J,L,-K_{\mathrm{sym}},Q_{\mathrm{sym}})=(32.5,65,180,1000)$~MeV.  
Results of the parameter survey for this set are shown in fig.~\ref{fig:jlkq}, where conventions are the same as in fig.~\ref{fig:skin}.
We will compare the results a)  from the above EoS and a resulting EDF including a modification parameter, shown with the solid black line in fig.~\ref{fig:EoSs} and b) from the EoS and resulting EDF with the same symmetry energy parameters but without the low-density modification, see dash-dotted line on fig.~\ref{fig:EoSs}.  
The former corresponds to a yellow point within the rectangle in fig.~\ref{fig:jlkq}, while the latter corresponds to one of the few black points (top right).
In each case the $D_{12}$ and $W_0$ parameters are again such that a good description of energies and charge radii is achieved in closed-shell nuclei. 
We note, in addition, that to reproduce the CREX and PREX-II results relatively high values are required. 
Here they equal $25$ and $140$~MeV~fm$^5$, respectively. Reconciliation can be achieved with lower values too, if a stiffer EoS is chosen.  
\begin{figure*}
\begin{center}
\includegraphics[width=15cm]{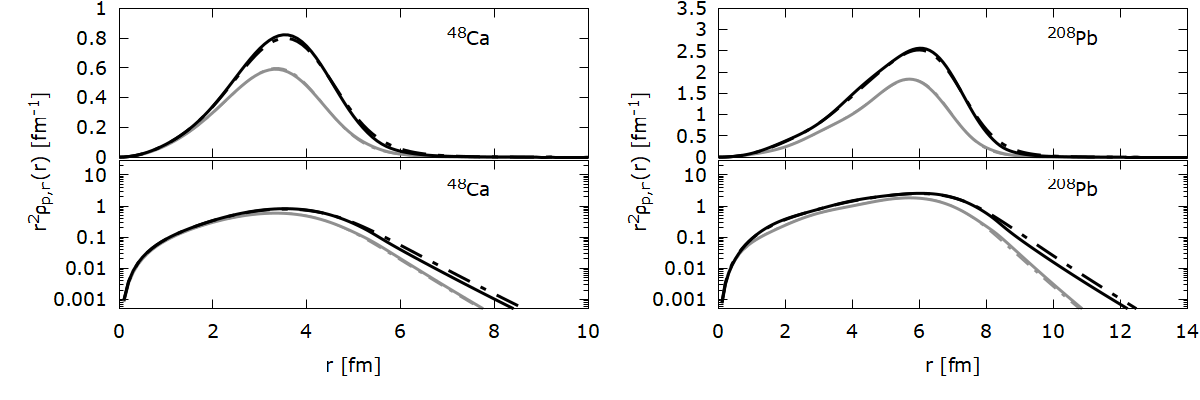}
\end{center} 
\vspace{-1cm}
\caption{Proton (gray lines) and neutron (black lines) density profiles of $^{48}$Ca (left) and $^{208}$Pb (right) in linear (top) and logarithmic scale. 
The dot-dashed lines, visible for neutrons, correspond to a regular KIDS EoS and EDF with the symmetry-energy parameters $(J,L,K_{\mathrm{sym}},Q_{\mathrm{sym,}}=(32.5,65,-180,1000)$~MeV. 
The full lines correspond to an EoS with the same standard parameters but including an enhancement at very low densities. 
The effect is a small change in the neutron-density diffuseness which suffices to reconcile the CREX and PREX-II values. 
\label{fig-example} }
\end{figure*} 
The corresponding density profiles for  $^{48}$Ca and $^{208}$Pb are shown in figure \ref{fig-example}. 
Here, it is seen that for the two EDF models corresponding to exactly the same EoS above $\rho_0/10$ and with the same quality in terms of energies and charge radii, the neutron density distribution can be much modified by the low-density effects and the neutron skin thickness can be much different as a result. 

Table~\ref{tab-example} 
shows the results for the neutron skin thickness and charge radii 
obtained for $^{48}$Ca, $^{120}$Sn and $^{208}$Pb with the two EDF models.  
As desired and by design, the charge radii are comparable for both functionals. % in all three nuclei.  
They all lie within about $1\%$ from experimental data. 
However, the neutron skin thickness is reduced substantially when the modification term is introduced---by an amount well exeeding the variations in charge (and proton) radii. 
Intrestingly, the neutron skin thickness of $^{120}$Sn is obtained similar to that of $^{48}$Ca. 

\begin{table}
\flushleft{%\centering
\caption{Representative results from two EDF models corresponding to exactly the same EoS above $\rho_0/10$ as discussed in the text. 
\label{tab-example} }       }% Give a unique label
\centering
% For LaTeX tables you can use
\begin{tabular}{rccc}
\hline \\[-3mm]
       & $^{48}$Ca  & $^{120}$Sn & $^{208}$Pb  \\ 
\hline \\[-3mm]
  $R_{np}$ [fm] KIDS plus:     & 0.171 & 0.172 & 0.212 \\ % & 0.170 & 0.165 & 0.213 \\
                           KIDS orig:      & 0.230 & 0.234 & 0.277 \\ %& 0.275 & 0.274 & 0.336  \\

\hline \\[-3mm]
  $R_{\mathrm{ch}}$ [fm] KIDS plus: & 3.507 & 4.624 & 5.463 \\ %& 3.511 & 4.618 & 5.447 \\
                                        KIDS orig:      & 3.497 & 4.608 & 5.450 \\ %& 3.491 & 4.598 & 5.441 \\
                                              exp.:        &  3.477 &  4.652  &  5.501 \\ 
\hline
\end{tabular}
% Or use
\vspace*{5cm}  % with the correct table height
\end{table}

\section{Conclusion} 

The PREX-CREX puzzle can be resolved within non-relativistic EDF theory and with realstic EoS parameters, such that no conflict arises with the properties of compact stars. This can be achieved by introducing an effective correction to the dilute matter EoS. %The higher-order regime, which is important for the description of neutron stars, is not affected. 
%On the other hand, t
The required adjustments to the gradient terms of the functional and the spin-orbit terms suggest that a further exploration of these properties is in order n particular in the isovectore channel. This work points to subtle structure and surface effects at play which are not captured by the EoS near the saturation regime. 
{\bf Acknowledgments} This work was supported by the Institute for Basic Science (2013M7A1A1075764).

%\bibliography{./biblio/bib-kids.bib,./biblio/bib-skin.bib}

\begin{thebibliography}{}
\bibitem{Oertel2017} M. Oertel, M. Hempel, T. Klähn, S. Typel, Rev. Mod.
Phys. 89, 015007 (2017).
\bibitem{Roca2018} X. Roca-Maza, N. Paar, Prog. Part. Nucl. Phys. 101,
96 (2018).
\bibitem{Garg2018} U. Garg, G. Colò, Prog. Part. Nucl. Phys. 101, 55
(2018).
\bibitem{ijmpe31}  H. Gil, P. Papakonstantinou, C.H. Hyun, Int. J. Mod.
Phys. E 31, 2250013 (2022).
\bibitem{XuP2022}  J. Xu, P. Papakonstantinou, Phys. Rev. C 105,
044305 (2022).
\bibitem{kids_nuclei1}  H. Gil, P. Papakonstantinou, C.H. Hyun, Y. Oh, Phys.
Rev. C 99, 064319 (2019).
\bibitem{PREX}  D. Adhikari et al. (PREX Collaboration), Phys. Rev.
Lett. 126, 172502 (2021).
\bibitem{CREX}  D. Adhikari et al. (CREX Collaboration), Phys. Rev.
Lett. 129, 042501 (2022).
\bibitem{rila2022}  P. Papakonstantinou (2022), arXiv:2210.02696,
Proc. 39th Int. Workshop on Nuclear Theory, Rila
Mountains, Bulgaria, July 3-9, 2022
\bibitem{Li2017}  B.A. Li, Nuclear Physics News 27, 7 (2017).
\bibitem{symmetry} P. Papakonstantinou, C.H. Hyun, Symmetry 15
(2023).
\bibitem{Nat2010}  J.B. Natowitz, G. Röpke, S. Typel, D. Blaschke,
A. Bonasera, K. Hagel, T. Klähn, S. Kowalski,
L. Qin, S. Shlomo et al., Phys. Rev. Lett. 104, 202501
(2010).
\bibitem{Wada2012}  R. Wada, K. Hagel, L. Qin, J.B. Natowitz, Y.G. Ma,
G. Röpke, S. Shlomo, A. Bonasera, S. Typel, Z. Chen
et al., Phys. Rev. C 85, 064618 (2012).
\end{thebibliography}

\newpage 

\clearpage

\end{document}